\documentclass[reprint, aps, pre]{revtex4-1}
\usepackage{soul}
\usepackage{amsmath}
\usepackage{amssymb} 
\usepackage{bm}
\usepackage{upgreek} 
\usepackage{graphicx}
\usepackage[version=4]{mhchem} 
\usepackage[colorlinks=true,citecolor=blue,linkcolor=blue]{hyperref} 

\begin{document}

\title{Compression and acceleration of ions by ultra-short ultra-intense azimuthally-polarized light}
\author{Da-Chao Deng}
\author{Hui-Chun Wu}
\email{huichunwu@zju.edu.cn}
\affiliation{Institute for Fusion Theory and Simulation, School of Physics, Zhejiang University, Hangzhou 310058, China}
\date{\today}
\begin{abstract}
  An efficient plasma compression scheme by azimuthally-polarized (AP) light is proposed. An AP light possesses a donut-like intensity pattern, enabling it to compress and accelerate ions toward the optical axis across a wide range of parameters. When the light intensity reaches the relativistic regime of $10^{18}$ $\mathrm{W}/\mathrm{cm}^{2}$, and the plasma density is below the critical density, protons can be compressed and accelerated by the toroidal soliton formed by the light. The expansion process of the soliton can be well described by the snow-plow model. Three-dimensional (3D) particle-in-cell (PIC) simulations show that within the soliton regime, despite the ion density surpassing ten times of the critical density, their energy is relatively low for efficient neutron production. When the light intensity increases to $10^{22}$ $\mathrm{W}/\mathrm{cm}^{2}$, and the plasma density is tens of the critical density, deuterium ions can be compressed to thousands of the critical density and meanwhile accelerated to the MeV level by a tightly-focused AP light during the hole-boring (HB) process. This process is far more dramatic compared to the soliton regime, and can produce up to $10^{4}$ neutrons in a few light cycles. Moreover, in the subsequent beam-target stage, neutron yield is assessed to reach over $10^{8}$. Finally, we present a comparison with the results by a radially-polarized (RP) light to examine the influence of light polarization.
\end{abstract}

\maketitle

\section{Introduction}
Neutron sources have broad applications in material testing for fusion energy research \cite{perkinsInvestigationHighIntensity2000}, neutron resonance spectroscopy \cite{higginsonLaserGeneratedNeutron2010}, radiation medicine \cite{schardtHeavyionTumorTherapy2010}, etc. With the development of laser technology, laser-driven neutron sources (LDNS) are attracting great attention due to its compactness, high transportability and ability for generating ultra-short neutron pulse \cite{alejoRecentAdvancesLaserdriven2016,yogoAdvancesLaserdrivenNeutron2023}. One of the currently investigated mechanisms capable of producing neutrons is nuclear-fusion reactions, suitable to be driven by laser owing to their relatively modest ion energy requirements compared to spallation reactions \cite{alejoRecentAdvancesLaserdriven2016}. In the LDNS schemes by fusion reactions, ion acceleration is necessary for optimizing the neutron yields, since the reaction cross peaks for the center-of-mass energy on the order of keV to MeV, depending on the specific reaction involved \cite{atzeniPhysicsInertialFusion2004}. There have been several LDNS schemes based on different ion acceleration mechanisms such as laser-cluster interactions by Coulomb explosion \cite{ditmireNuclearFusionExplosions1999}, in-target reactions by HB mechanism \cite{willingaleComparisonBulkPitchercatcher2011,toupinNeutronEmissionDeuterated2001} and beam-target fusion by target normal sheath acceleration \cite{willingaleComparisonBulkPitchercatcher2011,jiangEnergeticDeuteriumionBeams2020,yanInvestigationSphericallyConvergent2022}.

In order to enhance nuclear reactions within a limited time interval, some LDNS schemes employ the ion compression at the same time of their acceleration, such as spherically convergent plasma fusion \cite{zhuoLaserdrivenInertialIon2009,heSphericalShellTarget2015,renNeutronGenerationLaserDriven2017}, fusion enhanced by a double-cone configuration \cite{huLaserdrivenHighenergydensityDeuterium2015}, and the collapse of plasma shell with two ultra-intense laser pulses \cite{xuProductionHighdensityHightemperature2012}. In these schemes, ions can be compressed to $10$ to $10^{3}$ times the plasma critical density while accelerated to MeV, aided by specific target configurations or multiple laser pulses. As the reaction rate of nuclear fusion is proportional to the square of the ion density \cite{atzeniPhysicsInertialFusion2004}, it is possible to produce a significant number of neutrons in a very short period by compressing ions, which should be helpful for the generation of a short neutron pulse. Furthermore, the plasma compression via laser is a crucial concern for the exploit of inertial confinement fusion (ICF) energy. The current ion-compression schemes essentially possess structures similar to existing ICF schemes. There is a need for the exploration of novel plasma compression mechanisms, which can lead to the new LDNS schemes and also benefit the ICF research.

In this paper, we propose an efficient plasma compression mechanism, naturally achieved by the interaction an ultra-short ultra-intense AP light and uniform plasma, without the need for complex target configurations or multiple laser pulses. Recently, the focused laser intensity has surpassed $10^{23}$ $\mathrm{W}/\mathrm{cm}^{2}$ \cite{yoonRealizationLaserIntensity2021}, and the generation of few-cycle cylindrical vector (CV) light over $10^{19}$ $\mathrm{W}/\mathrm{cm}^{2}$ has been achieved through pulse compression \cite{kongGeneratingFewcycleRadially2019a} or the use of a optical converter \cite{carbajoEfficientGenerationUltraintense2014,carbajoDirectLongitudinalLaser2016}. It is important to explore the phenomena induced by CV light in the relativistic regime. Both AP and RP lights are CV-typed and share a cylindrically symmetric structure \cite{zhanCylindricalVectorBeams2009,forbesStructuredLightLasers2019}. Their amplitude shape is the same as that of the Laguerre-Gaussian (LG) mode with the radial index $p=0$ and the azimuthal index $l=1$, denoted as the $\mathrm{LG}_{01}$ mode \cite{zhanCylindricalVectorBeams2009}. Therefore, they typically exhibit a hollow intensity profile. When tightly focused, AP light generates a strong magnetic field along the optical axis on its focal point, whereas RP light gives a longitudinal electric field \cite{quabisFocusingLightTighter2000a,dornSharperFocusRadially2003}. In the relativistic regime, an AP light is discovered to form a long-standing toroidal soliton when penetrating through an underdense plasma \cite{chengRelativisticToroidalLight2023}. With a further increase in light intensity and plasma density, the HB process becomes to dominate the laser-plasma interaction \cite{yuEnhancedHoleBoring2016,psikalDominanceHoleboringRadiation2018}. We will show that ions can be compressed and accelerated to different levels by an AP light in these relativistic-soliton and HB regimes. It should be noted that the proposed plasma compression here does not work for a linearly- \cite{esirkepovThreeDimensionalRelativisticElectromagnetic2002,sarriObservationPostsolitonExpansion2010,toupinNeutronEmissionDeuterated2001} or circularly- \cite{liGenerationElectromagneticSolitons2021,wengUltraintenseLaserPulse2012} polarized light with a simple Gaussian profile under the similar laser-plasma interaction condition.

The rest of the paper is organised as follows. In Sec. \ref{sec2}, we present the compression phenomenon in the soliton regime by 3D PIC simulation. An analytic theory based on the snow-plow model is derived to describe the expansion of the toroidal soliton and plasma compression on its center. In Sec. \ref{sec3}, we investigate the compression and acceleration of deuterium ions in the HB regime and estimate the neutron yields from the reaction \ce{D + D -> ^3He + n}. Additionally, a comparison with RP light is presented. Discussions and conclusions are given in Sec. \ref{sec4}.

\section{The soliton regime}
\label{sec2}
\subsection{PIC simulation setup}
The PIC code used here is JPIC-3D \cite{wuJPICHowMake2011,chengRelativisticToroidalLight2023,chengRelativisticElectronBunches2023}, which adopts a field solver free of numerical dispersion along the $x$, $y$, $z$ axes. The AP light has an azimuthally-directed electric field $\bm{E}=E_{\theta}\bm{e}_{\theta}$ in the cylindrical coordinates $(r,\theta,x)$, and $E_{\theta}$ can be expressed as \cite{zhanCylindricalVectorBeams2009}
\begin{equation}
  \begin{split}
    E_{\theta} & =\sqrt{2\mathrm{e}}E_{0}r\frac{w_{0}}{w(x)^{2}}\exp\left[-\frac{r^{2}}{w(x)^{2}}\right]\\
    & \times f(t)\exp\left\{ i\left[kx-\omega t-2\phi(x)+\frac{kr^{2}}{2R(x)}\right]\right\},
  \end{split}
\end{equation}
where $\mathrm{e}$ denotes the natural number, $r=\sqrt{y^{2}+z^{2}}$ is the transverse radius, $w(x)=w_{0}\sqrt{1+x^{2}/X_{\mathrm{R}}^{2}}$ is the beam width, $\phi(x)=\arctan\left(x/X_{\mathrm{R}}\right)$ is the Gouy phase, $R(x)=x\left(1+X_{\mathrm{R}}^{2}/x^{2}\right)$ is the wavefront radius, $X_{\mathrm{R}}=kw_{0}^{2}/2$ is the Rayleigh length, $w_{0}$ is the beam waist at the focal point and $k=2\pi/\lambda$ is the wave number. The temporal profile takes $f(t)=\exp\left[-(t-x/c)^{2}/T_{0}^{2}\right]$, where $T_{0}$ is the pulse duration. The electric field amplitude $E_{0}$ can be normalized as $a_{0}=eE_{0}/m_{\mathrm{e}}\omega c$, where $\omega$ is the light angular frequency, $m_{\mathrm{e}}$ is electron mass, $e$ is electron charge, and $c$ is light speed. Laser intensity is given by $I_{0}[\mathrm{W}\big/\mathrm{cm}^{2}]\approx1.37\times10^{18}\lambda_{\upmu \mathrm{m}}^{-2}a_{0}^{2}$.

In the soliton regime, the laser pulse takes $a_{0}=1$, which corresponds to $I\approx1.37\times10^{18}$ $\mathrm{W}\big/\mathrm{cm}^{2}$ for the wavelength $\lambda=1$ $\upmu\mathrm{m}$. The laser has the spot size $w_{0}=3$, and the duration $T_{0}=2$, i.e. $6.7$ fs at $\lambda=1$ $\upmu\mathrm{m}$. The laser pulse propagates from the left boundary at $x=0$, and focuses on the plasma surface at $x=1\lambda$. The plasma length is $6\lambda$. The initial plasma density is $n_{\mathrm{e}}=n_{\mathrm{i}}=0.7n_{\mathrm{cr}}$, where $n_{\mathrm{cr}}=\epsilon_{0}m_{\mathrm{e}}\omega^{2}/e^{2}$ is the plasma critical density, and $\epsilon_{0}$ is the vacuum permittivity. The ions are chosen as protons with the mass $m_{\mathrm{i}}=1836m_{\mathrm{e}}$. The initial temperatures are $10$ eV for both electrons and ions.

The size of the simulation box is $8\lambda\times10.67\lambda\times10.67\lambda$, and the spatial resolution is 30 cells per wavelength, with each cell containing 8 macro-particles. The time step is $30^{-1}\lambda/c$.

\subsection{Compression and acceleration process of the plasma}
\begin{figure}[t]
  \centering
  \includegraphics[scale=0.55]{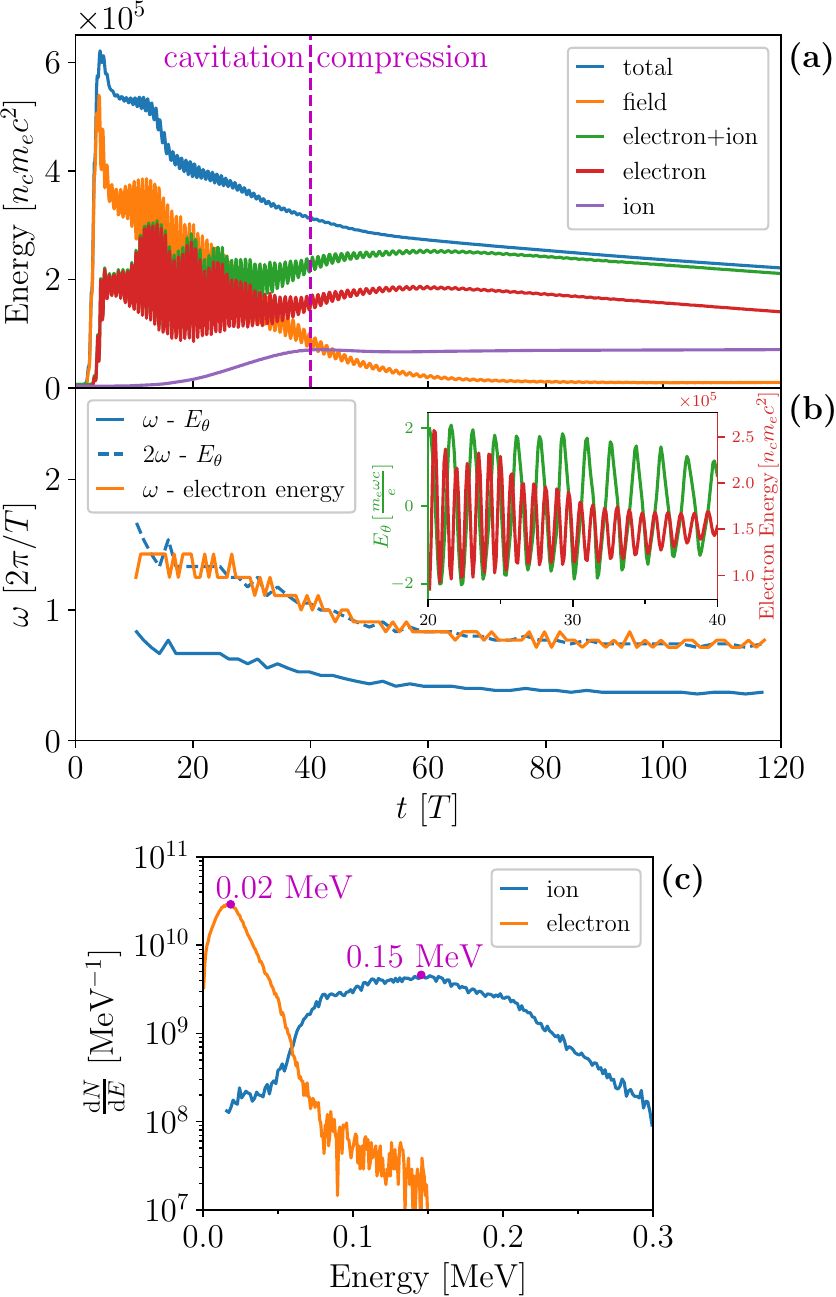}
  \caption{(a) Temporal evolution of the field and particles energy in the simulation box. The magenta dashed line at $t=40\lambda/c$ divides the cavitation and compression stages. (b) Oscillation frequency of the azimuthal electric field at $(x,y,z)=(4.3,6.8,5.3)\lambda$ and of the electron energy. The inset displays the temporal evolution of the azimuthal electric field and the electron energy from $t=20\lambda/c$ to $40\lambda/c$. (c) Energy spectra of the electrons and ions in the red rectangular zone in Fig. \ref{fig2}-(c). }
  \label{fig1}
\end{figure}
\begin{figure}[ht]
  \centering
  \includegraphics[scale=0.48]{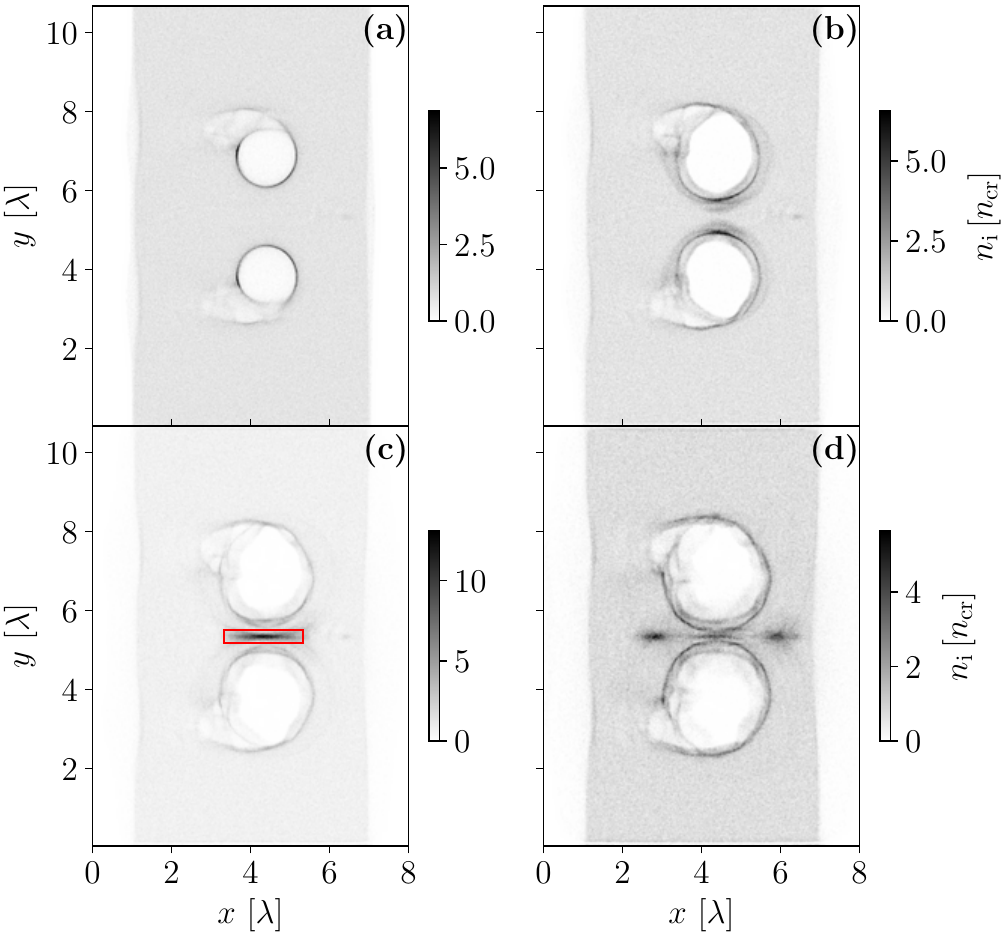}
  \caption{Compression stage of the soliton. Ion density distributions taken in the $x$-$y$ plane at $z=5.3\lambda$ for $ct/\lambda=40$ (a), $70$ (b), $89.3$ (c), $120$ (d).}
  \label{fig2}
\end{figure}
Fig. \ref{fig1}-(a) shows the energy evolution in the whole simulation box from $t=0$ to $120\lambda/c$. The whole process consists of two stages: the cavitation stage before $t=40\lambda/c$ and the compression stage thereafter. At the cavitation stage, the AP light has been fully input after $t=6\lambda/c$. The electrons are accelerated quickly with an increasing kinetic energy as they consume laser field energy. A fraction of light is reflected out of the simulation box, which leads to the first steep descent of the total energy. The second steep descent occurring at $t\approx16\lambda/c$ can be attributed to light transmission across the plasma out of the right boundary. Owing to the downshift in the light frequency below the background plasma frequency, a large part of light gets trapped in the plasma after $t=20\lambda/c$. This trapped light then pushes electrons outward by the ponderomotive force, forming a subcycle toroidal soliton as first discussed in Ref. \onlinecite{chengRelativisticToroidalLight2023}. The decline of the total energy starts to exhibit a more gentle slope since the trapped soliton slowly radiates out its energy. On a timescale of the ion response time $2\pi\omega_{\mathrm{pi}}^{-1}$, the ions initiate movement and cavitation in response to the space-charge field, marking the onset of the soliton's evolution into a postsoliton \cite{naumovaFormationElectromagneticPostsolitons2001,esirkepovThreeDimensionalRelativisticElectromagnetic2002,chengRelativisticToroidalLight2023}. At $t=40\lambda/c$ in Fig. \ref{fig2}-(a), the ion energy climbs up to a flatform, and here an ion toroidal cavity forms as wide as the electron one.

After $t=40\lambda/c$, as the ions gradually catch up the electrons, the whole toroidal plasma cavity enters into the compression stage. The toroidal postsoliton resembles a bent two-dimensional s-polarized postsoliton \cite{naumovaFormationElectromagneticPostsolitons2001,sanchez-arriagaTwodimensionalPolarizedSolitary2011,sanchez-arriagaTwodimensionalPolarizedSolitary2011a}. Therefore, ions on the inner side of the torus can be compressed and accelerated toward the torus center due to this distinct topological structure. Since the azimuthally-directed electric field is tangential to the boundary of the toroidal cavity, the ponderomotive force that the light exerts on the plasma is more stable with a weak field-particle energy exchange. As a result, the energy oscillation amplitude becomes small, and the subsequent compression is relatively moderate.

It should be noted that the ponderomotive force holds a crucial role in both the cavitation and compression stages. To see this, Fig. \ref{fig1}-(b) shows that the oscillation frequency of the azimuthal electric field is exactly twice that of the electron energy (the inset shows temporally). This frequency characteristic arises from the fact that the pondermotive force of this azimuthal soliton contains a second harmonic oscillation \cite{wengUltraintenseLaserPulse2012} $f_{p}=-\frac{1}{4}m_{\mathrm{e}}c^{2}\nabla'a^{2}(\bm{x}')[1+\cos(2\omega t)]$.

The later evolution of the compression stage is presented from Figs. \ref{fig2}-(b) to \ref{fig2}-(d). At $t=70\lambda/c$, the particles on the inner side of the torus start to merge and penetrate with each other. At $t=89.3\lambda/c$, they are compressed to a narrow density needle with a width of only around $0.6\lambda/c$. The ion density reaches $n_{\mathrm{i}}\simeq13n_\mathrm{cr}$, ten times of the initial density. Fig. \ref{fig1}-(c) gives the energy spectra of the particles in the red rectangular zone in Fig. \ref{fig2}-(c). Fitted with the Maxwellian distribution, the ion spectrum reveals an effective ion temperature on the order of 0.1 MeV. This ion energy is relatively low for the efficient neutron production \cite{atzeniPhysicsInertialFusion2004}. At $t=120\lambda/c$, the ion density needle is further squeezed into two parts, travelling towards the opposite directions along the torus axis.

\subsection{The snow-plow model}

\begin{figure}[htbp]
  \centering
  \includegraphics[scale=0.3]{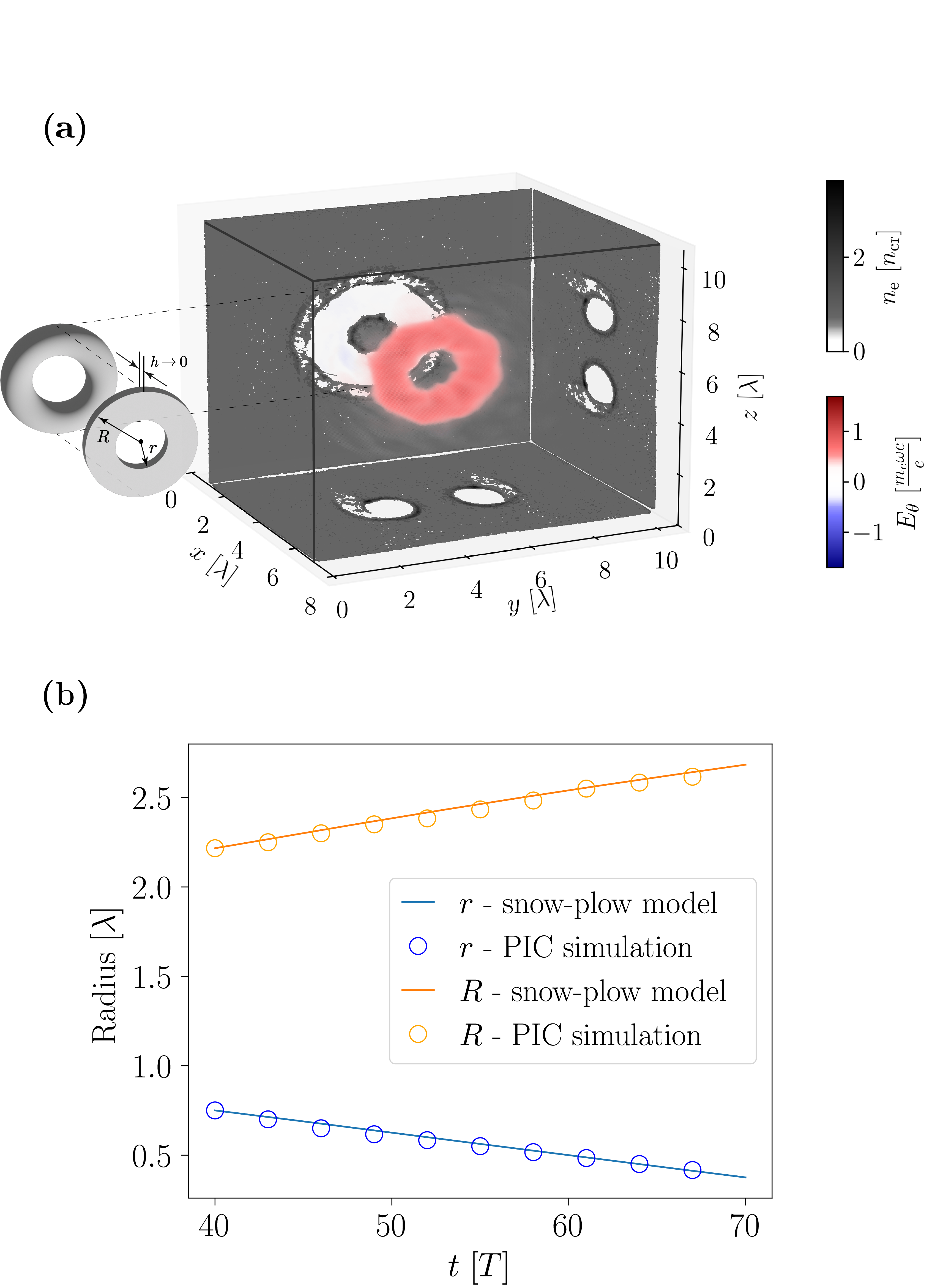}
  \caption{(a) 3D diagram of the azimuthal electric field at $t=40\lambda/c$. Projections in the $x$-$y$, $y$-$z$ and $z$-$x$ planes show the electron density taken at $z=5.3\lambda$, $x=4.3\lambda$ and $y=5.3\lambda$ respectively. The torus represents the plasma cavity, and the hollow cylinder is the approximation we use to model the enlargement of the annulus in the equatorial plane of the torus. $r$ is the inner radius, $R$ is the outer radius, and $h$ is the thickness of the hollow cylinder. (b) Comparison of the results from the snow-plow model and PIC simulation.}
  \label{fig3}
\end{figure}

The snow-plow model has been used to study the long-term expansion of two-dimensional postsolitons formed by s-polarized lights in Refs. \onlinecite{naumovaFormationElectromagneticPostsolitons2001,sanchez-arriagaTwodimensionalPolarizedSolitary2011,sanchez-arriagaTwodimensionalPolarizedSolitary2011a}. Here, we employ the snow-plow model to describe the expansion process of the toroidal cavity from $t=40\lambda/c$ to $70\lambda/c$. First, as shown in Fig. \ref{fig3}-(a), we use a flat hollow cylinder with an infinitesimal thickness $h$ to approximately represent the enlargement of the annulus in the equatorial plane. Here $r$ is its inner radius and $R$ the outer radius. We use the subscripts ``in'' and ``out'' to distinguish the quantities on the inner or outer side of the torus. The snow-plow model assumes that all the particles pushed by the light pressure $\simeq\epsilon_{0}\langle E^{2}\rangle/2$ are inside an thin shell. Let $\sigma$ be the ion surface number density in the cylinderical surface, and hence the plasma mass can be expressed as
\begin{equation}
  \begin{split}
    M_{\mathrm{in }} & =2\uppi rhm_{\mathrm{i}}\sigma_{\mathrm{in }},\\
    M_{\mathrm{out }} & =2\uppi Rhm_{\mathrm{i}}\sigma_{\mathrm{out }},
  \end{split}
  \label{eq=inner and outter mass}
\end{equation}
while noting that the electron mass has been neglected. By the particle number conservation, we have
\begin{equation}
  \begin{split}
    2\uppi rh\sigma_{\mathrm{in }} & =2\uppi r_{0}h\sigma_{\mathrm{in,0}}+\uppi(r_{0}^{2}-r^{2})hn_{0},\\
    2\uppi Rh\sigma_{\mathrm{out }} & =2\uppi R_{0}h\sigma_{\mathrm{out,0}}+\uppi(R^{2}-R_{0}^{2})hn_{0},
  \end{split}
  \label{eq=conservation of particles}
\end{equation}
where the subscript ``0'' denotes the initial quantities at $t=40\lambda/c$. Here we further assume that the cylinderical surface with radius of $(r_{0}+R_{0})/2$ is the interface between the inner and outer sides of the hollow cylinder, and then the total number of the ions on the inner side can be calculated as
\begin{equation}
  \begin{split}
    N_{\mathrm{in}}=N\uppi h & =2\uppi r_{0}h\sigma_{\mathrm{in,0}}+\uppi r_{0}^{2}hn_{0}\\
    & =\uppi R_{0}^{2}hn_{0}-2\uppi R_{0}h\sigma_{\mathrm{out,0}}\\
    & =\uppi\left(\frac{R_{0}+r_{0}}{2}\right)^{2}hn_{0},
  \end{split}
  \label{eq=N_in}
\end{equation}
where the defined term $N$ for convenience is given as
\begin{equation}
  N=n_{0}\left(\frac{R_{0}+r_{0}}{2}\right)^{2}\equiv n_{0}L_{0}^{2},
  \label{eq=coefficient N}
\end{equation}
here $L_{0}$ is the major radius of the torus. Substituting Eqs. \eqref{eq=N_in} and \eqref{eq=coefficient N} into Eq. \eqref{eq=conservation of particles}, we obtain
\begin{equation}
  \begin{split}
    \sigma_{\mathrm{in }} & =\frac{1}{2}\left(\frac{N}{r}-n_{0}r\right),\\
    \sigma_{\mathrm{out }} & =\frac{1}{2}\left(n_{0}R-\frac{N}{R}\right).
  \end{split}
  \label{eq=inner and outter surface density}
\end{equation}
Since only the initial conditions and the particle number conservation law are used here, one can see that the surface number density or plasma mass is simply independent of the dynamical process in the snow-plow model.

For the electric field, we can relate it to the cavity geometry by introducing the adiabatic invariant \cite{naumovaFormationElectromagneticPostsolitons2001,landauElectrodynamicsContinuousMedia1984}
\begin{equation}
  \int_{}^ {}\frac{E^{2}}{\Omega}\mathrm{d}V
\end{equation}
and by approximating the toroidal postsoliton as a resonant mode \cite{capToroidalResonatorsElectromagnetic1978} $\Omega\sim\rho^{-1}=(R-r)^{-1}$. Thus these two equations give
\begin{equation}
  \langle E^{2}\rangle\simeq\langle E_{0}^{2}\rangle\left(\frac{R_{0}-r_{0}}{R-r}\right)^{3}.
  \label{eq=approximation of E}
\end{equation}

Finally, the dynamical equations driven by the radiation pressure can be written as
\begin{equation}
  \begin{split}
    \frac{\mathrm{d}}{\mathrm{d}t}\left(M_{\mathrm{in}}\frac{\mathrm{d}r}{\mathrm{d}t}\right) & =-2\uppi rh\frac{\epsilon_{0}\langle E^{2}\rangle}{2},\\
    \frac{\mathrm{d}}{\mathrm{d}t}\left(M_{\mathrm{out}}\frac{\mathrm{d}R}{\mathrm{d}t}\right) & =2\uppi Rh\frac{\epsilon_{0}\langle E^{2}\rangle}{2},
  \end{split}
  \label{eq=dynamics eqs}
\end{equation}
where $M$ is the plasma mass. We then substitute Eqs. \eqref{eq=inner and outter mass}, \eqref{eq=inner and outter surface density} and \eqref{eq=approximation of E} to Eqs. \eqref{eq=dynamics eqs}, and obtain the evolution equations of $r$ and $R$ as
\begin{equation}
  \begin{split}
    (N-n_{0}r^{2})\frac{\mathrm{d}^{2}r}{\mathrm{d}t^{2}}-2 & n_{0}r\left(\frac{\mathrm{d}r}{\mathrm{d}t}\right)^{2}\\
    & =-\frac{\epsilon_0 r}{m_{\mathrm{i}}}\langle E_{0}^{2}\rangle\left(\frac{R_{0}-r_{0}}{R-r}\right)^{3},\\
    (n_{0}R^{2}-N)\frac{\mathrm{d}^{2}R}{\mathrm{d}t^{2}}+2 & n_{0}R\left(\frac{\mathrm{d}R}{\mathrm{d}t}\right)^{2}\\
    & =\frac{\epsilon_0 R}{m_{\mathrm{i}}}\langle E_{0}^{2}\rangle\left(\frac{R_{0}-r_{0}}{R-r}\right)^{3}.
  \end{split}
  \label{eq=dynamics eqs(PIC units) of hollow cylinder model}
\end{equation}
These equations are solved numerically, and Fig. \ref{fig3}-(b) shows that the numerical solutions are in very good agreement with the PIC results.

\section{The hole-boring regime}
\label{sec3}
\begin{figure*}[t]
  \centering
  \includegraphics[scale=0.42]{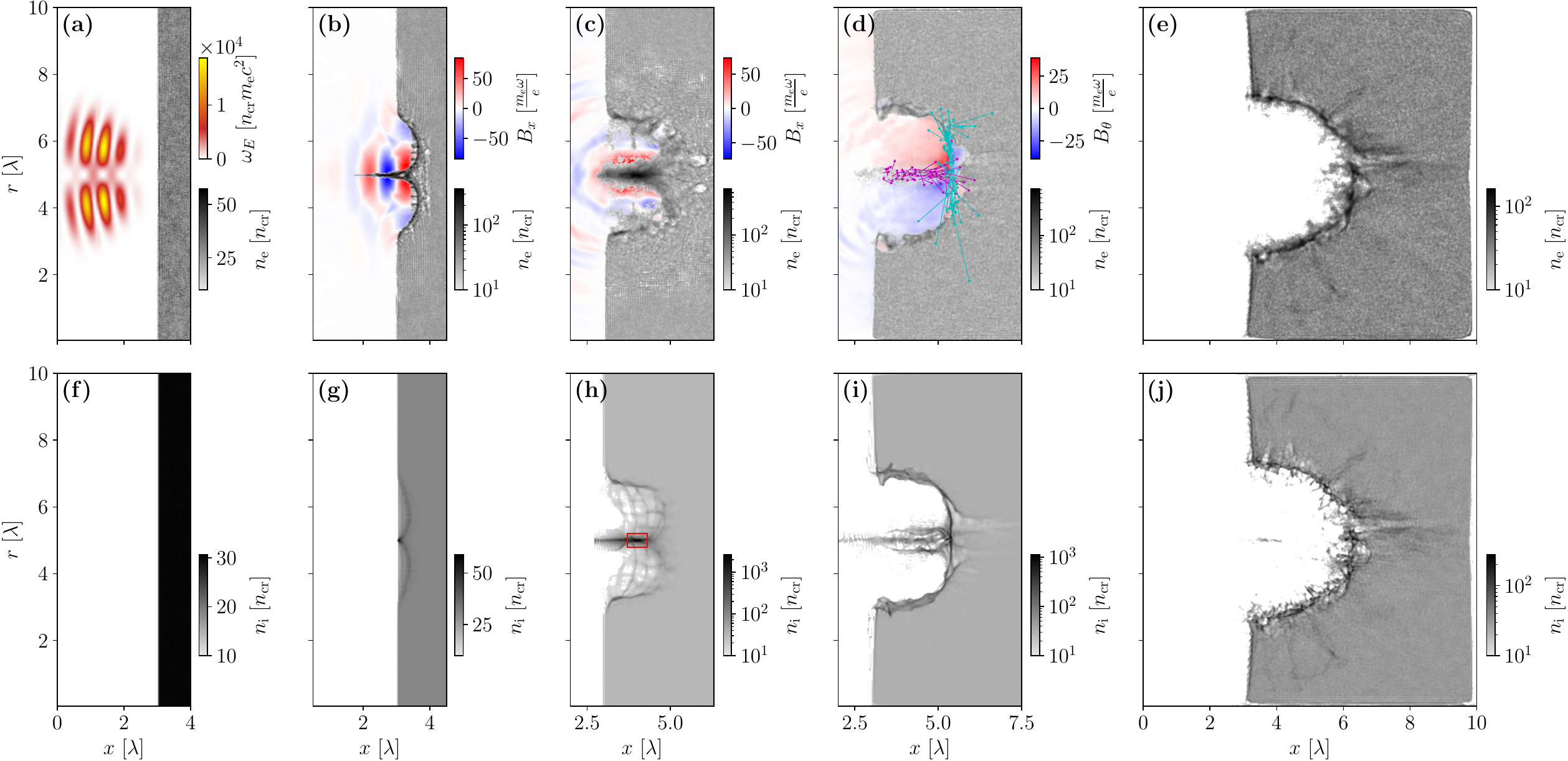}
  \caption{HB compression process by the AP light. Electron density (a)-(e) and ion density (f)-(j) at $ct/\lambda=4$ (a)(f), $6$ (b)(g), $10.6$ (c)(h), $16.8$ (d)(i) and $40$ (e)(j). The field energy density, longitudinal magnetic field and azimuthal magnetic field are also shown in (a), (b)-(c) and (d), respectively. The magenta arrows in (d) represent the longitudinal current near the optical axis, while the cyan arrows represent the radial currents at the bottom of the plasma dip. }
  \label{fig4}
\end{figure*}

\subsection{Azimuthally-polarized light}
In this section, we attempt to enhance the plasma compression efficiency for a potential LDNS by raising light intensity. We will explore the so-called hole-boring regime where AP light can directly compress and accelerate ions. The normalized light amplitude takes $a=150$, which corresponds to $I\approx3.08\times10^{22}$ $\mathrm{W}\big/\mathrm{cm}^{2}$. The laser focus spot is shrunk to $w_{0}=1\lambda$, which can make the compression process faster and more efficient. When tightly focused to the wavelength scale, the AP light can generate a longitudinal magnetic field as strong as the transverse field \cite{quabisFocusingLightTighter2000a,dornSharperFocusRadially2003}. This field is helpful to confine the compressed electrons on the optical axis. In JPIC-3D, the laser-boundary-condition algorithm \cite{thieleBoundaryConditionsArbitrarily2016} is adopted to produce an ideal focus field \cite{chengRelativisticElectronBunches2023}. The light is input at the left boundary and the focal plane is located at $x=3\lambda$. The plasma has the initial density $n_{0}=30n_\mathrm{cr}$. The ion mass is changed to $m_{\mathrm{i}}=3672m_\mathrm{e}$ to simulate a deuterium plasma. The size of the simulation box is $12\lambda\times10\lambda\times10\lambda$. All the other parameters remain the same as the soliton regime.

\begin{figure}[htbp]
  \centering
  \includegraphics[scale=0.405]{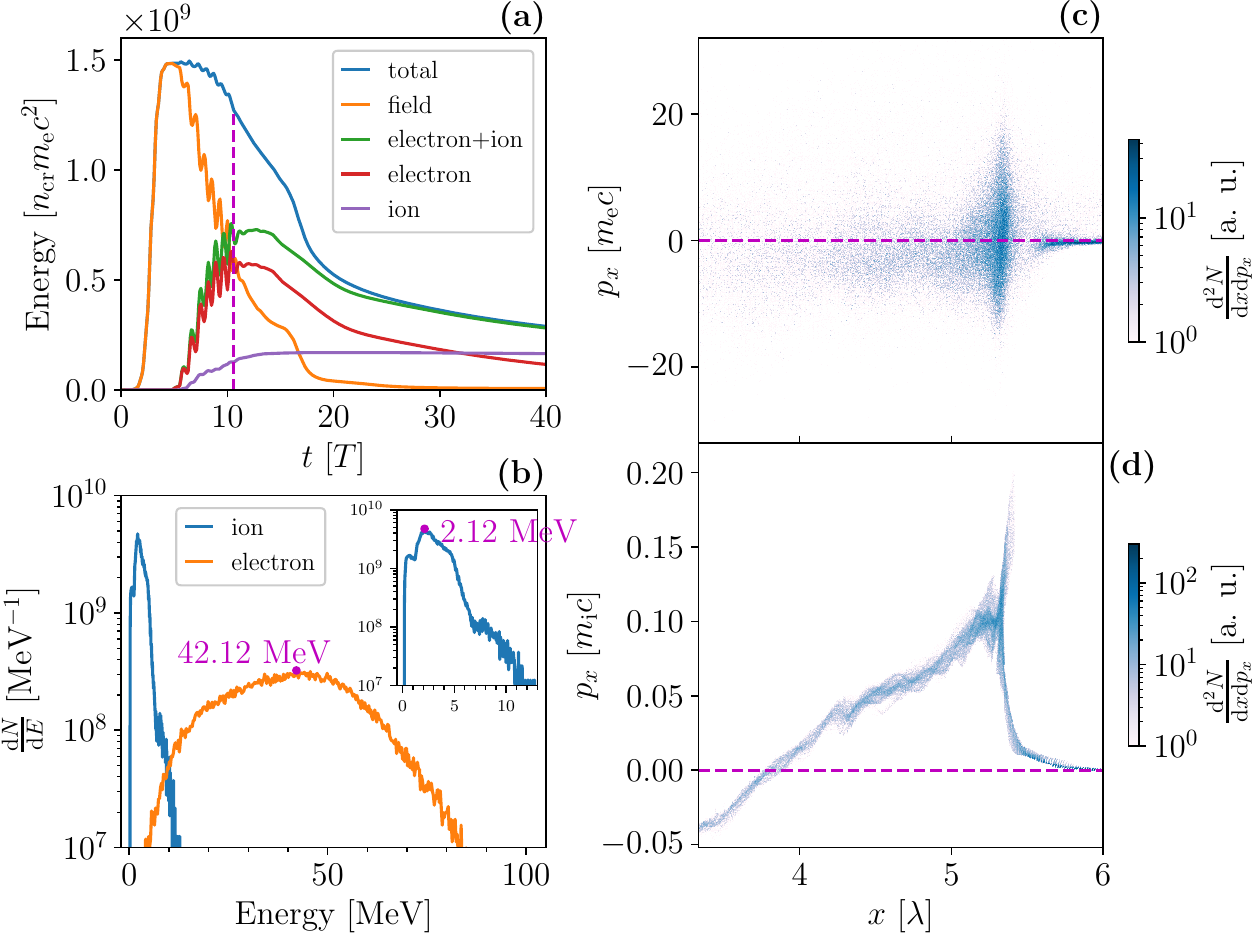}
  \caption{HB dynamics by the AP light. (a) Temporal evolution of field and particle energy in the simulation box. The magenta dashed line marks $t=10.6\lambda/c$. (b) Energy spectra of electrons and ions within the red rectangular zone in Fig. \ref{fig4}-(h). (c)-(d) Longitudinal momentum $p_{x}$ vs. position $x$ for electrons and ions located in the zone covered by the magenta arrows in Fig. \ref{fig4}-(d).}
  \label{fig5}
\end{figure}

\begin{figure*}[ht]
  \centering
  \includegraphics[scale=0.42]{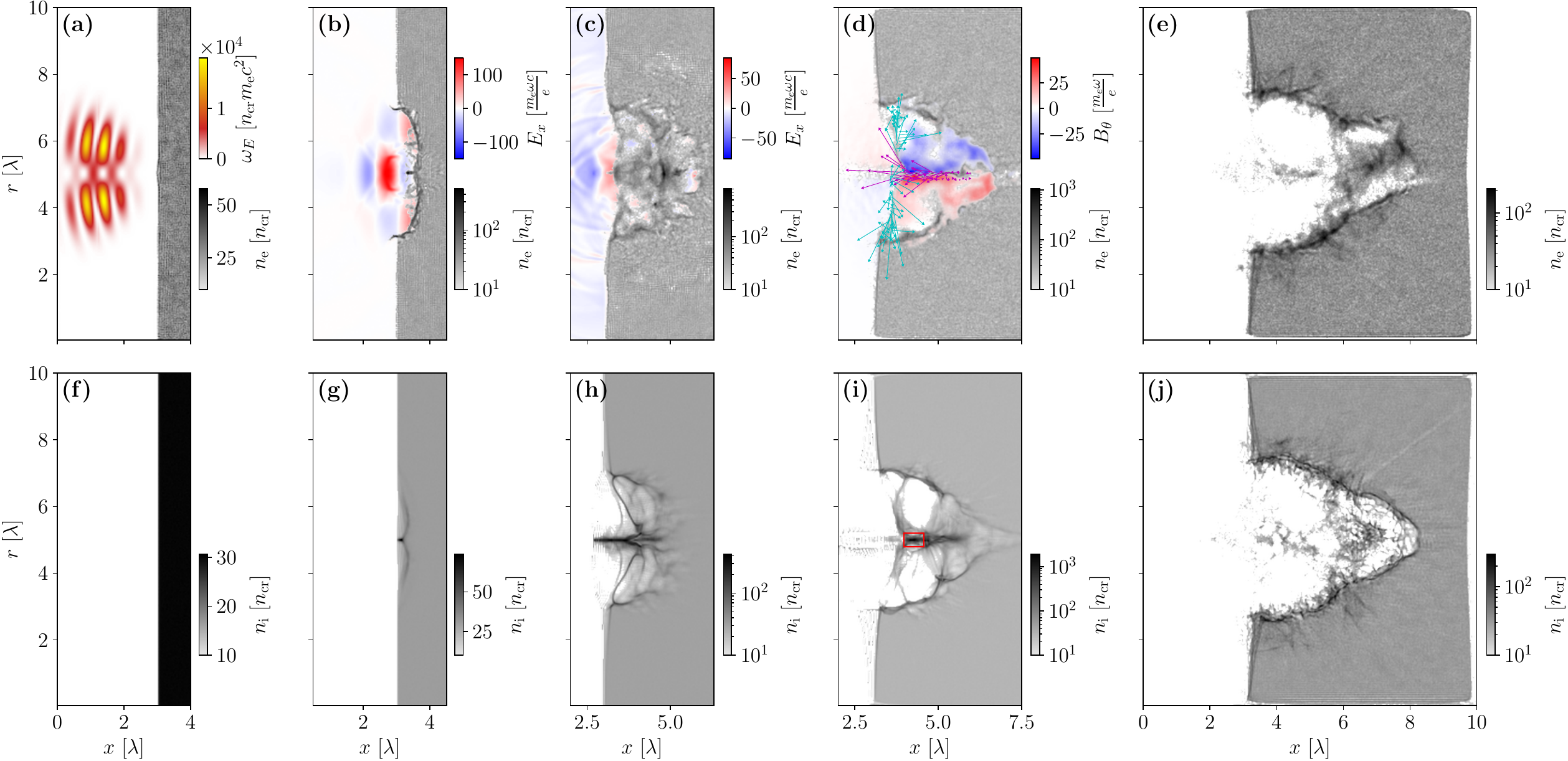}
  \caption{HB compression process by the RP light. Same as Fig. \ref{fig4} with two exceptions: first, it is longitudinal electric fields, rather than magnetic fields, that are shown in (b)-(c); second, radial currents are located at the top, rather than bottom, of the plasma dip in (d).}
  \label{fig6}
\end{figure*}

Fig. \ref{fig4} illustrates the plasma compression process by the AP light. In Fig. \ref{fig4}-(a), the tightly-focused AP light exhibits a hollow intensity pattern, which punches out a ring-like dip on the plasma surface through the light pressure at $t=6\lambda/c$, as shown in Fig. \ref{fig4}-(b) and \ref{fig4}-(g). The electrons on the optical axis are then squeezed backward and confined as a sharp needle due to the ponderomotive force and strong longitudinal magnetic field of the tightly-focused AP light. The ions lag behind, following the movement of the electrons due to their space-charge field. Fig. \ref{fig5}-(a) illustrates the energy evolution of the system from $t=0$ to $40\lambda/c$. It can be observed that at $t=6\lambda/c$, discernible oscillations in both field and electron energies manifest the role of the ponderomotive force.

At $t=10.6\lambda/c$, it is shown in Fig. \ref{fig4}-(c) and \ref{fig4}-(h) that the ions undergo compression, forming a ring-like dip comparable in width to that of the electrons and achieving the ion peak density of $n_{i}\gtrsim2600n_\mathrm{cr}$ on the optical axis. This ion compression is mainly attributed to the electro-static attraction exerted by the electrons compressed and confined on the optical axis. At this moment, as in Fig. \ref{fig5}-(a), the energy oscillation tends to diminish, indicating that the light will soon be completely reflected from the plasma region. Fig. \ref{fig5}-(b) presents the energy spectra of the particles within the red rectangular zone in Fig. \ref{fig4}-(h), revealing that the effective ion temperature reaches the order of 1 MeV, and electrons are $\sim 10$ MeV. The high density and energy of the ions are attained at the same time, providing a suitable environment for producing neutrons by nuclear reactions. From the neutron yield \cite{atzeniPhysicsInertialFusion2004} $Y=\frac{n_{\mathrm{D}}^{2}}{2}\langle\sigma_{\mathrm{D-D}}v\rangle V\Delta t$, where $n_{\mathrm{D}}$ is the deuterium density, $\langle\sigma_{\mathrm{D-D}}v\rangle$ is the average reactivity of the deuterium-deuterium reaction, $V$ is the volume of the reaction region, and $\Delta t$ is the reaction time, the neutron yield is estimated to be around $1.3\times10^{4}$ from $t=8\lambda/c$ to $12\lambda/c$, spanning around 13 fs.

At $t=16.8\lambda/c$ in Figs. \ref{fig4}-(d) and \ref{fig4}-(i), both compressed ions and electrons are further accelerated forward along the $+x$ direction and disperse slightly as the light departs, resulting in a decrease in the ion peak density to $n_{i}\approx1000n_\mathrm{cr}$. The relatively high ion density is maintained by the strong azimuthal quasi-static magnetic field generated by the longitudinal current in the compressed plasma column. As presented in Figs. \ref{fig5}-(c) and \ref{fig5}-(d), the ions moving forward and a fraction of electrons moving backward that dominate this positive longitudinal current. The presence of the radial currents is also observed at the bottom of the dip in Fig. \ref{fig4}-(d). In Fig. \ref{fig4}-(i), the finger-like plasma channels are also discernible within the plasma due to the penetration of the light as a result of relativistically induced transparency \cite{pukhovLaserHoleBoring1997}.

In final, the HB by the AP light leaves a U-shaped crater 3$\lambda$ in depth and 2.4$\lambda$ in diameter at $t=40\lambda/c$ in Figs. \ref{fig4}-(e) and \ref{fig4}-(j). A significant portion of piled-up and accelerated ions would penetrate and travel through the deuterium plasma, giving rise to a substantial neutron yield via the beam-target approach. The neutron yield by this approach can be estimated using the integral \cite{yogoAdvancesLaserdrivenNeutron2023,toupinNeutronEmissionDeuterated2001,heSphericalShellTarget2015}
\begin{equation}
  Y=n_{\mathrm{D}}\int_{0}^{\infty}\mathrm{d}E_{\mathrm{i}}f(E_{\mathrm{i}})\int_{0}^{E_{\mathrm{i}}}\mathrm{d}E\frac{\sigma(E)}{\varepsilon(E)},
\end{equation}
where $\sigma(E)$ is the reaction section and $\varepsilon(E)$ is the stopping power of the deuterium ion in the deuterium plasma. In our case, the neutron yield resulting from the interaction of the injected deuterium ions and the deuterium plasma is approximately $1.16\times10^{8}$, much higher than the yield in the compression stage owing to a sufficient reaction time.

\subsection{Radially-polarized light}

\begin{figure}[htbp]
  \centering
  \includegraphics[scale=0.41]{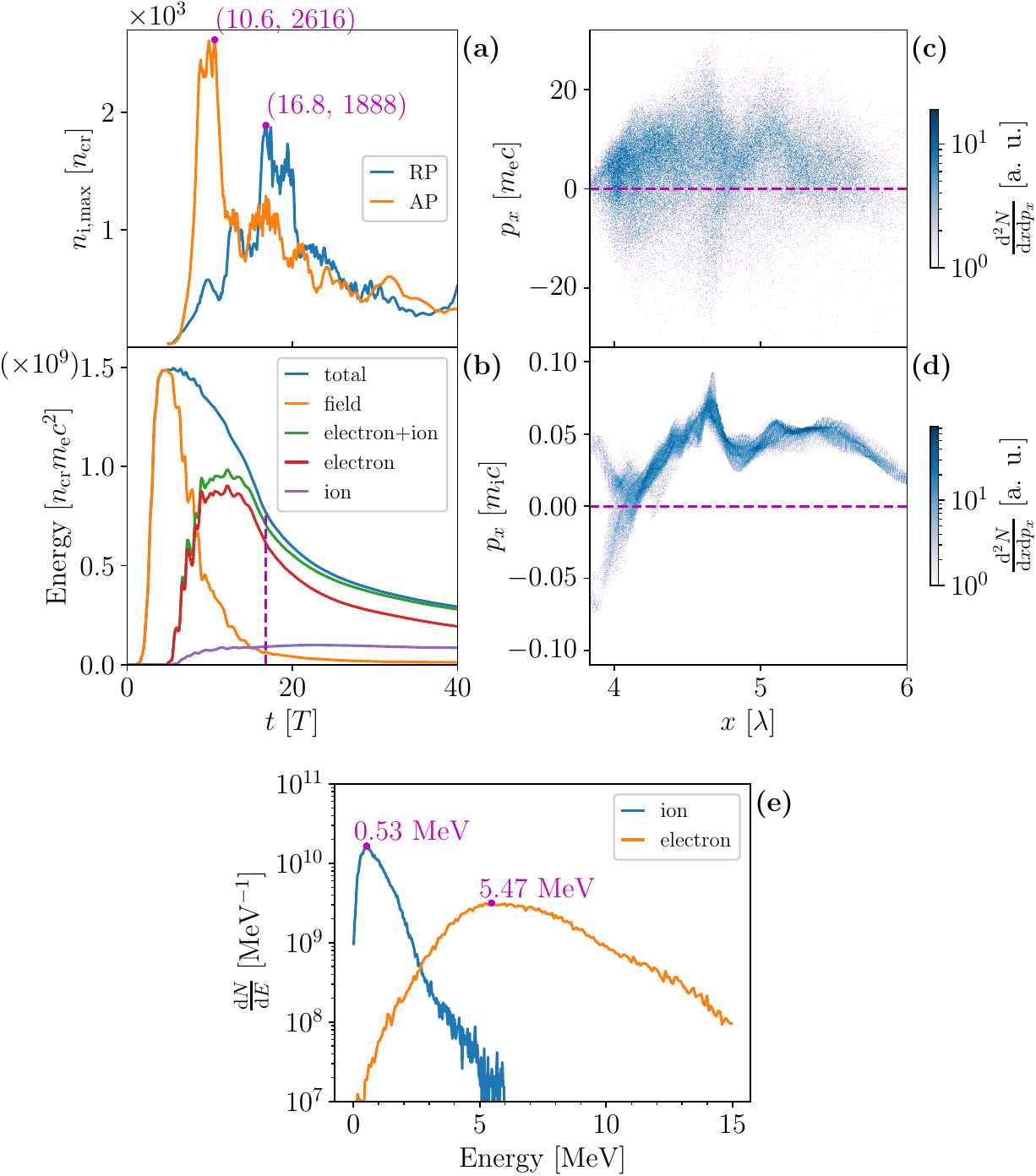}
  \caption{HB dynamics by the RP light. (a) Temporal evolution of the maximum ion density achieved by AP or RP light near the optical axis. (b) Temporal evolution of field and particle energy in the simulation box. The magenta dashed line marks $t=16.8\lambda/c$. (c)-(d) $p_{x}$ vs. $x$ for particles located in the zone covered by the magenta arrows in Fig. \ref{fig6}-(d). (e) Energy spectra of electrons and ions within the red rectangular zone in Fig. \ref{fig6}-(i). }
  \label{fig7}
\end{figure}

AP light has been shown to be a robust natural mode for forming a long-standing toroidal soliton with a relatively moderate plasma compression, and the change in polarization can influence its geometry and stability \cite{liSelforganizedFractallikeBehaviors2020}. However, the HB regime discussed above seems to be mainly dominated by the light ponderomotive force, which is simply proportional to the gradient of intensity distribution \cite{mourouOpticsRelativisticRegime2006}. To check the potential polarization effects in the HB regime, a case using RP light is displayed in Fig. \ref{fig6} and Fig. \ref{fig7}, with the other parameters fixed. The expression of a RP light can be derived directly from that of an AP light through the duality transformation \cite{aprilUltrashortStronglyFocused2010} $\bm{E}\rightarrow c\bm{B}$ and $\bm{B}\rightarrow-\bm{E}/c$. Thus, both lights share the same intensity pattern \cite{dornSharperFocusRadially2003} as can be verified by comparing Fig. \ref{fig4}-(a) and Fig. \ref{fig6}-(a). Due to the alignment of the electric field and the magnetic field is exchanged for both-typed lights, a strong longitudinal electric field, rather than a magnetic field, appears when the RP light is tightly focused.

The polarization change has an impact on the compression and acceleration of ions. Fig. \ref{fig7}-(a) presents the temporal evolution of the maximum ion density near the optical axis for both polarization states. There are two observations can be made: firstly, the ion peak density achieved by the RP light is 28\% lower than the AP case; secondly, the time $t=16.8\lambda/c$ at which the ion density reaches its peak for the RP light is delayed than that for the AP light. In fact, as shown in Fig. \ref{fig7}-(b), there is no oscillation in the field or electron energy at $t=16.8\lambda/c$, which means that the light has been completely scattered away from the plasma region.

Fig. \ref{fig6}-(g) illustrates the ion motion just during the reflection of the RP light. As shown in Fig. \ref{fig6}-(b), the electrons on the optical axis are pushed into the plasma rather than squeezed out in Fig. \ref{fig4}-(b). As the light penetrates deeper, the electrons are pulled back by the reversed longitudinal electric field in the second subcycle. This push-pull action \cite{zaimRelativisticAccelerationElectrons2017} shakes the electrons longitudinally, along with radially-oscillating motion induced by the radial electric field in the interaction region. As a result, the electrons can not be stably confined on the optical axis, leading to a lower ion density.

When the light leaves off the plasma region at $t=16.8\lambda/c$, as shown in Figs. \ref{fig6}-(d) and \ref{fig6}-(i), a negative longitudinal current is formed, generating a strong quasi-static azimuthal magnetic field. As seen in Figs. \ref{fig7}-(c) and \ref{fig7}-(d), the majority of electrons accelerated by its longitudinal electric field possess a positive longitudinal momentum, leading to this negative longitudinal current. This resulting azimuthal magnetic field pinches the electron current into a filament \cite{naumovaPolarizationHosingLong2001}, subsequently draws the ions toward the axis, analogous to the Z-pinch process \cite{slutzPulsedpowerdrivenCylindricalLiner2010,hurricanePhysicsPrinciplesInertial2023}. In addition, the radial current induced by the electrons moving from the radial boundary of the plasma to the optical axis can be found at the top of the dip, as shown in Figure \ref{fig7}-(d), resulting in the reversed azimuthal magnetic field at the top of the plasma dip.

The energy spectra of the particles within the red rectangular zone in Fig. \ref{fig6}-(i), when the ion density reaching its peak, is presented in Fig. \ref{fig7}-(e). The effective ion and electron temperature are only about 0.1 MeV and 1 MeV, respectively. Both are nearly one order of magnitude lower than that of the AP case. This is mainly because most energetic electrons accelerated by the push-pull mechanism \cite{zaimRelativisticAccelerationElectrons2017} have run out of the compression region as the light departs, and hence the energy transferred to ions is low, as discussed above.

The neutron yield by nuclear reaction from $t=15.8\lambda/c$ to $19.8\lambda/c$ is estimated to be $9.3\times10^{3}$, and the subsequent beam-target yield is estimated to be $2.38\times10^{7}$. Both are lower than the AP case. Nevertheless, owing to the strong longitudinal electric field of the RP light, the crater in Figs. \ref{fig6}-(e) and \ref{fig6}-(j) exhibits a width of 1.3 $\lambda$ and a significantly greater depth of 5.5 $\lambda$ when compared to that punched by the AP light.

\section{Discussion and conclusion}

\label{sec4}
\begin{figure}[htbp]
  \centering
  \includegraphics[scale=0.545]{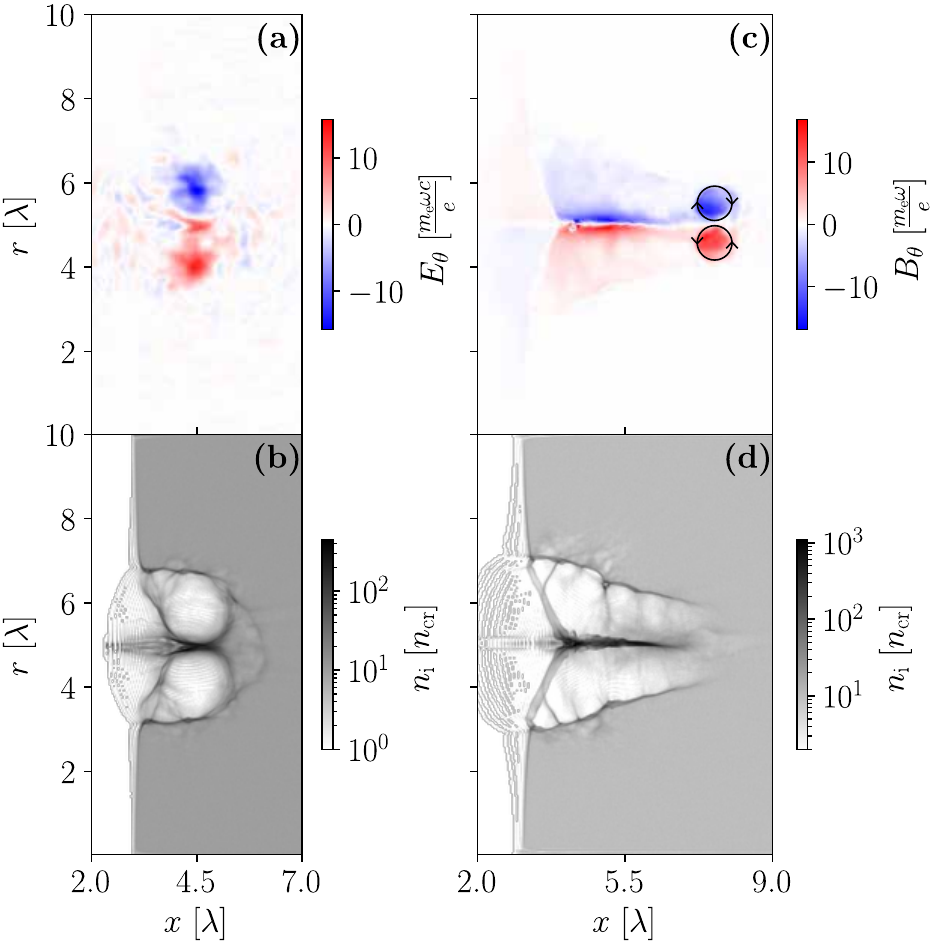}
  \caption{Intermediate case for $a_{0}=50$ and $n_{0}=10n_\mathrm{cr}$, and all other parameters are the same as Fig. \ref{fig4}. AP light: azimuthal electric field (a) and ion density (b) at $t=18.8\lambda/c$. RP light: azimuthal magnetic field (c) and ion density (d) at $t=25.9\lambda/c$. The black arrows in (c) point the direction of the electric currents.}
  \label{fig8}
\end{figure}

In the HB regime, a wide parameter range is scanned by changing light intensity and initial density. We find that, to achieve the optimum ion compression with the highest density, a positive correlation remains between light intensity and plasma density. If the plasma is too dense or the light is too weak, AP light is not able to drill deep enough to compress ions. Conversely, if the plasma is too rarefied or the light is too strong, the number of the ions involved in the compression is not large enough compared to that the light could compress.

Interestingly, it is noteworthy that even within the HB regime, AP light can be trapped in the ring-like plasma dip for several light cycles, behaving like a short-term soliton, as shown in Figs. \ref{fig8}-(a) and \ref{fig8}-(b). This case takes a relatively low light amplitude $a_{0}=50$ for a plasma density $n_{0}=10n_\mathrm{cr}$. The case using RP light is also shown in Figs. \ref{fig8}-(b) and \ref{fig8}-(d). The deuterium peak densities reach $n_{\mathrm{i}}\approx500n_\mathrm{cr}$ and $1000n_\mathrm{cr}$ for AP and RP lights, respectively. Additionally, Fig. \ref{fig8}-(c) illustrates that vortex electric currents can be generated, giving rise to a robust quasi-static azimuthal magnetic field, similar to that observed by using a Gaussian-shape light \cite{yueStructureTransportationElectron2021}. The RP light in this case can finally penetrate through the plasma, and release its energy at the plasma-vacuum interface, which has the potential to accelerate and collimate ions as discussed in Refs. \onlinecite{bulanovCommentCollimatedMultiMeV2007,nakamuraHighEnergyIonsNearCritical2010,helleLaserAcceleratedIonsShockCompressed2016}.

In terms of the polarization effect, we primarily focus on these two CV lights, as they have a cylindrically symmetric structure, ensuring a more uniform and stable compression and acceleration of ions and facilitating the analysis in both theory and simulation. On the other hand, $\mathrm{LG}_{01}$ light of linear or circular polarization has also demonstrated the potential to form a toroidal ion cavity in plasma \cite{wilsonLaserPulseCompression2019,wilsonSelffocusingCompressionCollapse2023} or compress ions toward the optical axis \cite{wangHollowScrewlikeDrill2015}. However, to the best of our knowledge, there has been no detailed discussion on the ion compression effect by ultrashort $\mathrm{LG}_{01}$ lights of these polarizations in the soliton or HB regime. Since these polarization states can be considered as a linear superposition of azimuthal and radial components in the cylindrical polarization bases \cite{ossikovskiPolarizedLightMueller2016,zhanPropertiesCircularlyPolarized2006}, our investigation on these two fundamental CV lights can serve as a foundation for further research concerned with polarization or other degrees of freedom in light, such as angular momentum \cite{zhanPropertiesCircularlyPolarized2006}.

In conclusion, our 3D PIC simulations show that the ion compression and acceleration can be realized efficiently by an AP light across a wide range of parameters. In the soliton regime, an AP light with the intensity of $10^{18}$ $\mathrm{W}/\mathrm{cm}^{2}$ can from a toroidal soliton and compress ions from $0.7n_\mathrm{cr}$ to $13n_\mathrm{cr}$, over ten times of the initial density. However, the ion energy is only on the 0.1 MeV level, not efficient for the neutron production. Based on the snow-plow model and some approximations, an analytic theory is derived to demonstrate the expansion of the toroidal cavity. In the HB regime, an AP light with the intensity level of $10^{22}$ $\mathrm{W}/\mathrm{cm}^{2}$ is adopted to drill into a deuterium plasma with initial density of $n_{0}=30n_\mathrm{cr}$. The peak deuterium ion density can surpass $2600n_\mathrm{cr}$, nearly one hundred times the initial density, and the ion energy can attain the MeV level. This efficient compression and acceleration of ions are appropriate for nuclear fusions. Here we consider the deuterium-deuterium reaction to estimate the neutron yields in different stages. In the compression stage, the neutron yield is $1.3\times10^{4}$ in 13 fs. The compressed ions can subsequently trigger nuclear fusions by the beam-target reactions as they penetrate through the deuterium plasma. In this stage, the neutron yield is around $10^{8}$, much higher than the yield in the compression stage due to a sufficient reaction time. To highlight the light polarization effect, ion compression by RP light is also examined. The dominant compression process by RP light resembles a Z pinch.

\begin{acknowledgments}
  D.-C. Deng would like to thank Dr. Y. Zhang and Dr. J.-L. Jiao for helpful discussions. This work was supported by the Strategic Priority Research Program of Chinese Academy of Sciences (Grant No.XDA17040502).
\end{acknowledgments}

\bibliography{Refs}

\end{document}